\documentclass[letterpaper, 10 pt, conference]{ieeeconf}  
\usepackage{amsmath}
\usepackage{amssymb}
\usepackage{amsfonts}
\usepackage{flushend}
\usepackage{mathrsfs}
\usepackage{graphicx}
\usepackage{psfrag}
\usepackage{slashbox}
\usepackage[linesnumbered,ruled]{algorithm2e}
\usepackage{url}
\usepackage{theorem}

\IEEEoverridecommandlockouts                              

\title{\LARGE \bf
Data-driven Thermal Model Inference with ARMAX, in Smart Environments, based on Normalized Mutual Information
}

\author{\begin{tabular}{cccc}
\small Zhanhong Jiang$^\dag$  & \small Jonathan Francis$^\ddag$$^\S$ & \small Anit Kumar Sahu$^\P$ & \small Sirajum Munir$^\S$\\
\small \tt zhjiang@iastate.edu  & \small \tt jmf1@cs.cmu.edu & \small \tt anits@ece.cmu.edu & \small \tt sirajum.munir@us.bosch.com
\end{tabular}\\ \vspace{3pt}
\begin{tabular}{ccc}
\small Charles Shelton$^\S$ & \small Anthony Rowe$^\P$ & \small Mario Berg\'es$^\diamond$\\
\small \tt charles.shelton@us.bosch.com & \small \tt agr@ece.cmu.edu& \small \tt marioberges@cmu.edu
\end{tabular}\\ \vspace{3pt}
\small $\dag$ Department of Mechanical Engineering, Iowa State University, Ames, IA 50011\\
\small $\ddag$ School of Computer Science, Carnegie Mellon University, Pittsburgh, PA 15213 \\
\small $\P$ Department of Electrical \& Computer Engineering, Carnegie Mellon University, Pittsburgh, PA 15213\\
\small $\S$ Bosch Research \& Technology Center North America, Pittsburgh, PA 15222\\
\small $\diamond$ Department of Civil \& Environmental Engineering, Carnegie Mellon University, Pittsburgh, PA 15213
}

\begin{document}

\maketitle
\thispagestyle{empty}
\pagestyle{empty}

\begin{abstract}

Understanding the models that characterize the thermal dynamics in a smart building is important for the comfort of its occupants and for its energy optimization. A significant amount of research has attempted to utlilize thermodynamics (physical) models for smart building control, but these approaches remain challenging due to the stochastic nature of the intermittent environmental disturbances. This paper presents a novel {\it data-driven} approach for indoor thermal model inference, which combines an Autoregressive Moving Average with eXogenous inputs model (ARMAX) with a Normalized Mutual Information scheme (NMI). Based on this information-theoretic method, NMI, causal dependencies between the indoor temperature and exogenous inputs are explicitly obtained as a guideline for the ARMAX model to find the dominating inputs. For validation, we use three datasets based on building energy systems|against which we compare our method to an autoregressive model with exogenous inputs (ARX), a regularized ARMAX model, and state-space models. 

\end{abstract}

\section{INTRODUCTION}

Maintaining occupant thermal comfort in buildings is key for facilitating productivity and health of those occupants and for ensuring efficient building facilities management; poor indoor thermal comfort can lead to less productivity, even some physiological and psychological problems~\cite{enescu2017review}. Maintaining thermal comfort in a building typically requires satisfying having sufficiently-adaptive environmental models to capture the dynamics of the zone(s) under consideration and to adjust for unforseen perturbations.
Two types of models are typically used to address the model-inference issue for air-temperature inside a building; firstly, thermodynamics models based on the physical dynamics of the integral-differential equations, with respect to energy and mass; and, secondly, data-driven models based on time-series measures of environmental sensor modalities (temperature, pressure, humidity, etc.) and the related  machine learning inference techniques. Technically speaking, whereas physical models are able to describe the indoor thermal dynamics accurately, deriving such models is sometimes not feasible when indoor environments are sophisticated and intermittently disturbed and they require more computational resources. Therefore, data-driven methodologies~\cite{kalogirou2000applications,rios2007modelling,chinde2015comparative} have recently received considerable attention and are shown to be quite effective at describing the indoor thermal conditions.

While data-driven approaches need not consider complex {\it physical} relationships among variables, they still have an issue in differentiating dominating inputs from disturbances. Moreover, among control inputs, one or two of these modalities may not have a significant impact on the indoor temperature, whereas considering all of them may significantly increase the computational resources required. Therefore, we assert that the causal dependencies between indoor temperature, the control inputs, and the disturbances should first be discussed, in order to establish an appropriate data-driven model. 

\textbf{Contributions}: First, we use ARMAX model to establish the indoor thermal dynamics and to discuss two scenarios, namely, single-input single-output (SISO) and multi-input single-output (MISO). We also compare ARMAX to ARX, regularized ARAMX, and state-space models. Second, according to results obtained from SISO and MISO, an information-theoretic metric framework based on normalized mutual information (NMI) is established; this allows us to quantify the causal dependencies between the indoor temperature and other variables for determining dominating exogenous inputs and intermittent disturbances, in order to provide a guideline for model inference. Finally, three different data sets on indoor thermal comfort are used to validate the proposed ARMAX scheme based on NMI.

\textbf{Related Work}: ANN and its variants have been shown useful in predicting the evolution of system dynamics and approximating highly nonlinear functions~\cite{marvuglia2014coupling}. In~\cite{liang2005thermal}, the authors used ANN to describe the indoor thermal comfort model for environment temperature regulation. ANN was also used to quantify the thermal behavior of a building based on a simulated program~\cite{kalogirou2000applications}, to enhance indoor thermal comfort and building energy efficiency~\cite{moon2016algorithm}, and to develop an advanced thermal control method for maintaining thermal comfort requirements in buildings~\cite{moon2010ann}. Furthermore, ARX and ARMAX~\cite{mustafaraj2010development} were applied for identifying the thermal behavior in office buildings. In~\cite{wu2012physics}, the authors presented a physics-based ARMAX to predict the room temperature in both short-term and long-term periods. A neural network-based nonlinear ARX (NNARX)~\cite{mechaqrane2004comparison} was proposed and developed to predict the room temperature and relative humidity of an open office.

\section{Indoor Thermal Dynamics by ARMAX}\label{ARMAX}

This section presents the typical indoor thermal temperature dynamics of a room by ARMAX and shows initial results based on SISO and MISO. We first give some preliminaries about ARMAX models.
\subsection{ARMAX}
A generic ARMAX model~\cite{mustafaraj2010development} can be described as follows
\begin{equation}\label{armax}
\Phi(z)y_t=\Lambda(z)x_{t-\alpha} + \Theta(z)\epsilon_t,
\end{equation}
where $x$ is the external input variable, $y$ is the output variable, $\epsilon$ is the white noise representing intermittent disturbances, $\alpha$ is the system time delay between input and output, and $z$ indicates the backshift operator. Therefore, we have
\begin{subequations}
\begin{alignat}{3}\label{backshift_operator}
  &\Phi(z) = 1+\phi_1z^{-1}+\phi_2z^{-2}+\dots+\phi_{n\phi} z^{-n\phi} \\
  &\Lambda(z) = 1+\lambda_1 z^{-1}+\lambda_2 z^{-2}+\dots+\lambda_{n\lambda} z^{-n\lambda} \\
  &\Theta(z) = 1+\theta_1 z^{-1}+\theta_2 z^{-2}+\dots+\theta_{n\theta} z^{-n\theta},
\end{alignat}
\end{subequations}
where $n\phi,n\lambda,n\theta$ are the highest orders of backshift operators, $\Phi,\Lambda,\Theta$, respectively. From the above equation, it can be observed that, for ARMAX, the critical problem is to estimate coefficients associated with the polynomials $\Phi, \Lambda, \Theta$. However, external inputs may include many variables|some of which may have relatively small impacts on the output variable $y$. Equivalently, those variables cannot significantly improve the model-predictive capability; instead, they may be regarded as disturbances. Similarly, some variables that are regarded as disturbances can have significant effect on indoor thermal comfort, such as occupancy. 
\subsection{Indoor Thermal Dynamics: SISO and MISO}
Typically the indoor temperature is affected by multiple different factors, e.g., natural ventilation, supply air, lighting, occupancy, equipment, humidity, etc. 
For motivation, we next discuss the SISO and MISO cases based on a real data set~\cite{candanedo2016accurate}.
Description of the data set is as follows: there are 6 variables, including occupancy (binary mode, 0 or 1), temperature (degrees Celsius), relative humidity (\%), light (lux), $CO_2$ (ppm), and humidity ratio (HR, derived from temperature and relative humidity, in kg water-vapor/kg-air). One training dataset 
and two testing datasets 
were used; the test bed is an office room with approximate dimensions of 5.85 m$\times$3.50 m$\times$3.53 m ($W\times D\times H$) in Mons, Belgium and kept above $19^\circ$C in the winter season.

For the rest of paper, we evaluate the accruacy of one method based on the normalized root-mean-square error (NRMSE) between the testing data output and model prediction. We term NRMSE as model fit and the formula is as follows:
\begin{equation}
model\;fit = 100\%(1-\frac{\|y-\hat{y}\|_2}{\|y-\bar{y}\|_2}),
\end{equation}
where $y$ is the testing data output, $\hat{y}$ represents the model prediction, $\bar{y}$ is the mean of the testing data output, and $\|\cdot\|_2$ is the Euclidean norm.
\subsubsection{SISO}
\begin{figure}
  \centering
  \vspace{3pt}
  \includegraphics[width=0.3\textwidth,trim=.5cm .5cm .5cm .5cm,clip]{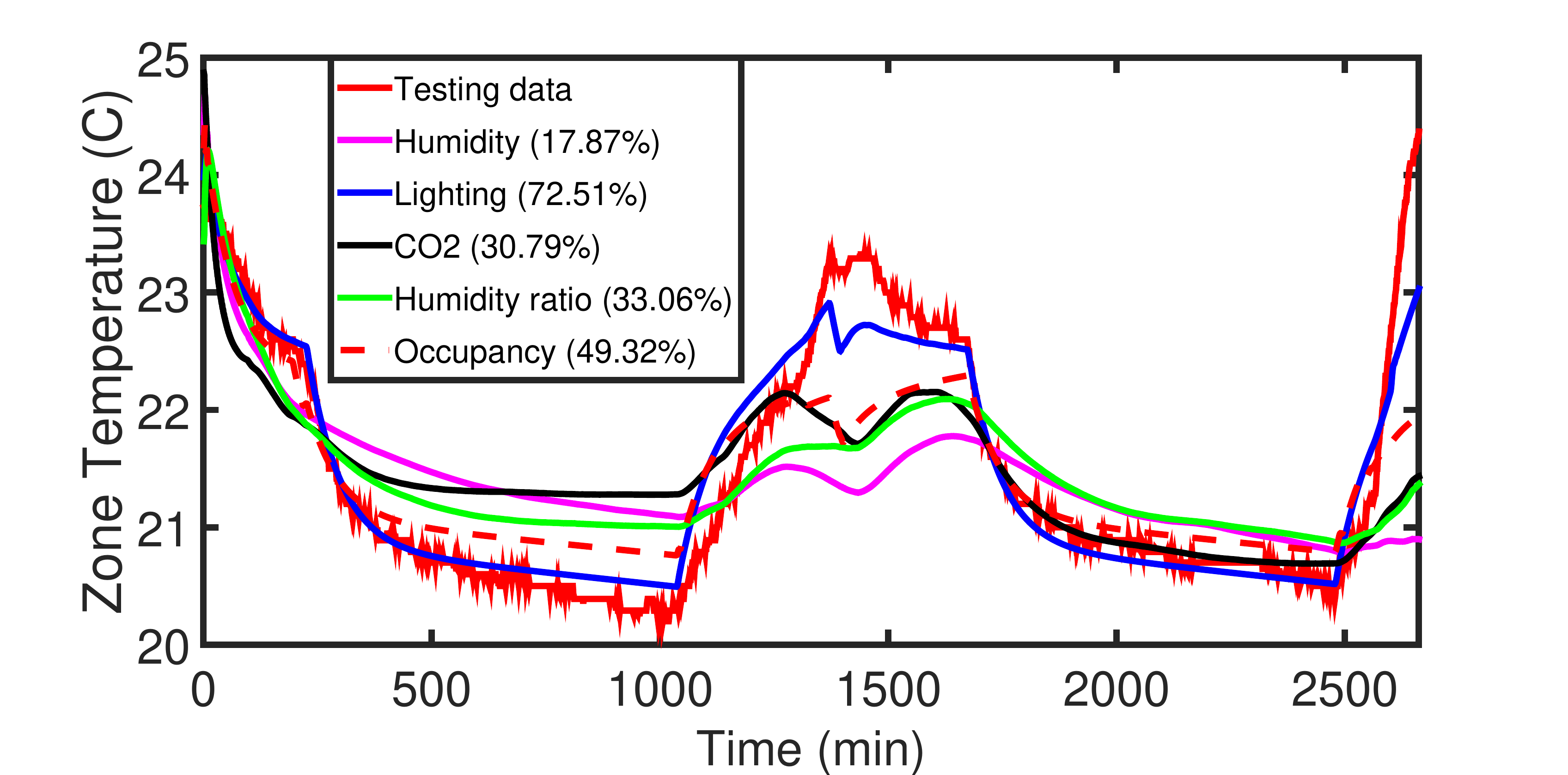}\vspace{-8pt}
  \caption{\textit{SISO with the first testing data set}}\label{Figure2}
\end{figure}

\begin{figure}
  \centering
  \includegraphics[width=0.3\textwidth]{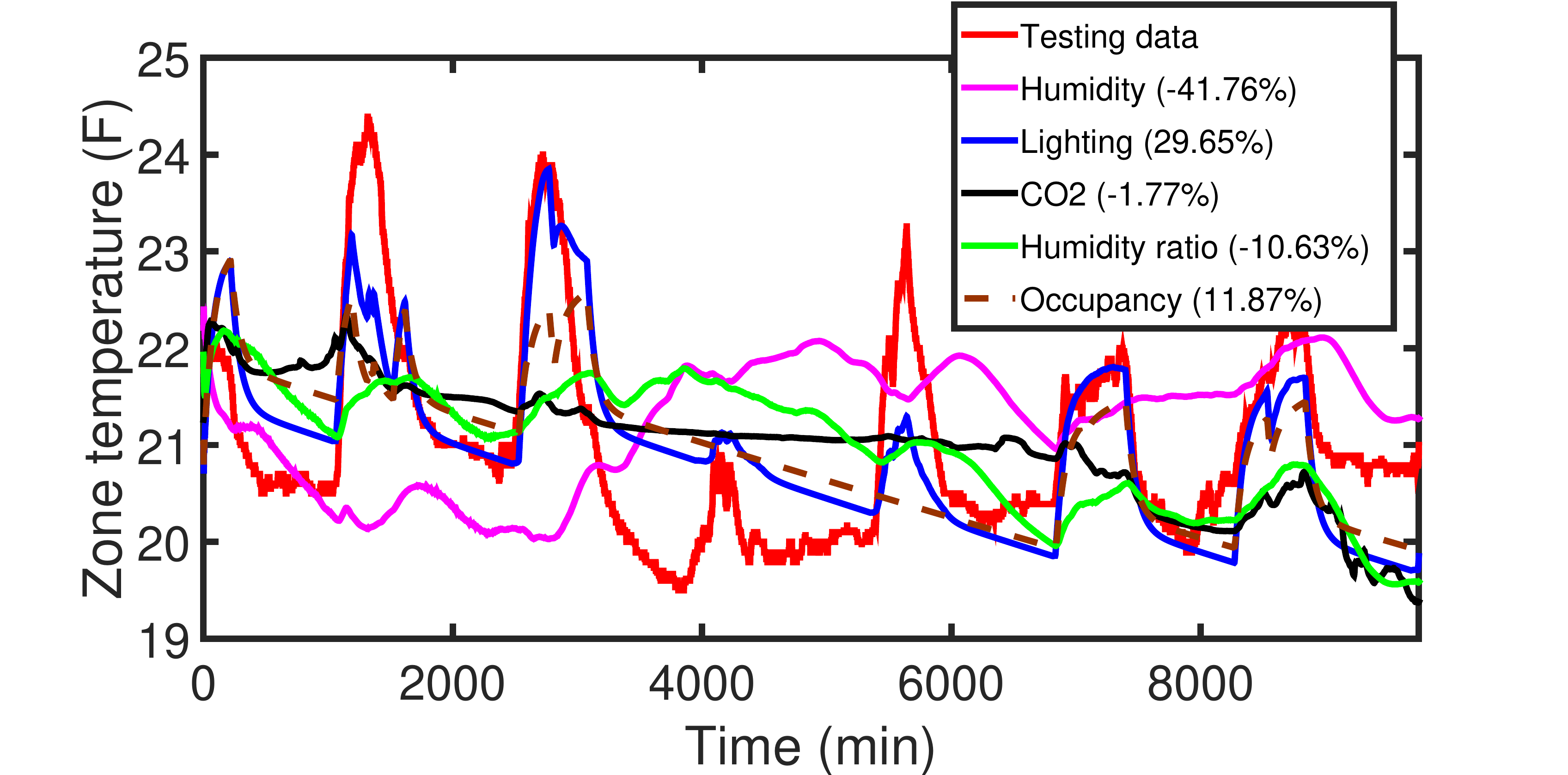}\vspace{-8pt}
  \caption{\textit{SISO with the second testing data set}}\label{Figure3}
\end{figure}

As the indoor temperature is the unique output, input is one of the other variables (occupancy, relative humidity, light, $CO_2$, or humidity ratio) and the remaining four variables are regarded as disturbances (white noise). Based on the SISO form of Eq.~\ref{armax}, fourth-order with delay set as 1 in this context, results on predictive capability are obtained in Figures~\ref{Figure2} and~\ref{Figure3}.
As shown in Figure~\ref{Figure2}, it can be observed that lighting is the dominating factor that has the most significant effect on the indoor air temperature, with 72.51\% model fit. Occupancy also has a relatively significant impact on the indoor air temperature, with 49.32\% model fit. 
Figure~\ref{Figure3} also forms a similar conclusion, where the predictive capability with only one control input is worse than that in Figure~\ref{Figure2}. 
In this dataset, some key variables, such as supply air and supply air static pressure (which are generally used as inputs in controller design) are not included. However, in terms of data-driven approaches, exogenous inputs can be uncontrolled, which motivates us to further consider MISO.
\subsubsection{MISO}
This section presents indoor thermal dynamics based on ARMAX with multiple exogenous inputs. While not every input combination is discussed, we combine lighting and occupancy with other variables to respectively check two and three inputs. The goal is to see if multiple exogenous inputs can improve the predicting capability of indoor thermal temperature.

Table~\ref{Table1} shows model fit results using multiple inputs. It can be observed that for the first testing data set, multiple inputs (lighting + occupancy) slightly improves the model predicting capability, while for the second testing data set, multiple inputs (Occupancy + $CO_2$) has worse prediction compared to the unique input (Occupancy). Thus, not all input combinations can be used for performance-improvement, and it turns out that the indoor air temperature in this case significantly depends on lighting, which equivalently suggests that the indoor thermal dynamics pattern can be reflected mostly by lighting pattern. Table~\ref{Table1} also shows the indoor air temperature prediction, using 3 exogenous inputs (Lighting + $CO_2$ + Occupancy) for two testing datasets. Results imply that for the first testing dataset, 3 inputs slightly improve the model's prediction capability. However, for the second testing dataset, the performance improvement is 13.60\%. This may be attributed to different inputs contributing to model accuracy in different time intervals.

\begin{table*}
\centering
\vspace{5pt}
\caption{Model fit in MISO cases}
\begin{tabular}{|l|*{3}{c|}}\hline
\backslashbox{Data set}{Variables}
&Lighting+Occupancy&$CO_2$+Occupancy&Lighting+$CO_2$+Occupancy
\\\hline
First testing data set &73.09\%&51.39\%&73.96\%\\\hline
Second testing data set &29.86\%&2.27\%&43.46\%\\\hline
\end{tabular}
\label{Table1}
\end{table*}

\section{Input Selection, Using NMI}\label{NMI}

Section~\ref{ARMAX} presents SISO and MISO cases which investigate the effect of different numbers of exogenous inputs on the output variable, based on a real dataset. However, there is no general approach to decide which variables are dominating exogenous inputs and which are noise, for the indoor thermal temperature models. To address this issue, we use an information-theoretic metric, based on normalized mutual information (NMI) to quantify causal dependencies between different variables. In this context, we consider the pair-wise variables and focus on the causal relationships between the indoor air temperature and other variables|called relational patterns~\cite{jiang2017energy}. Before introducing NMI, we first briefly present some preliminary discussion of mutual information (MI) for completeness.

A generic information-theoretic metric based on mutual information between two discrete random variables $U$ and $V$ can be defined as:
\begin{equation}\label{MI}
I(U;V) = \sum_{v\in V}\sum_{u\in U}p(u,v)\textnormal{log}\bigg(\frac{p(u,v)}{p(u)p(v)}\bigg),
\end{equation}
where $p(u,v)$ represents the joint probability distribution function of $X$ and $Y$, and $p(u)$ and $p(v)$ are the marginal probability distribution functions of $U$ and $V$, respectively. For more details about how to derive MI from the information entropy, we refer interested readers to~\cite{jiang2017energy}. 

\begin{figure}
  \centering
  \vspace{3pt}
  \includegraphics[width=0.3\textwidth]{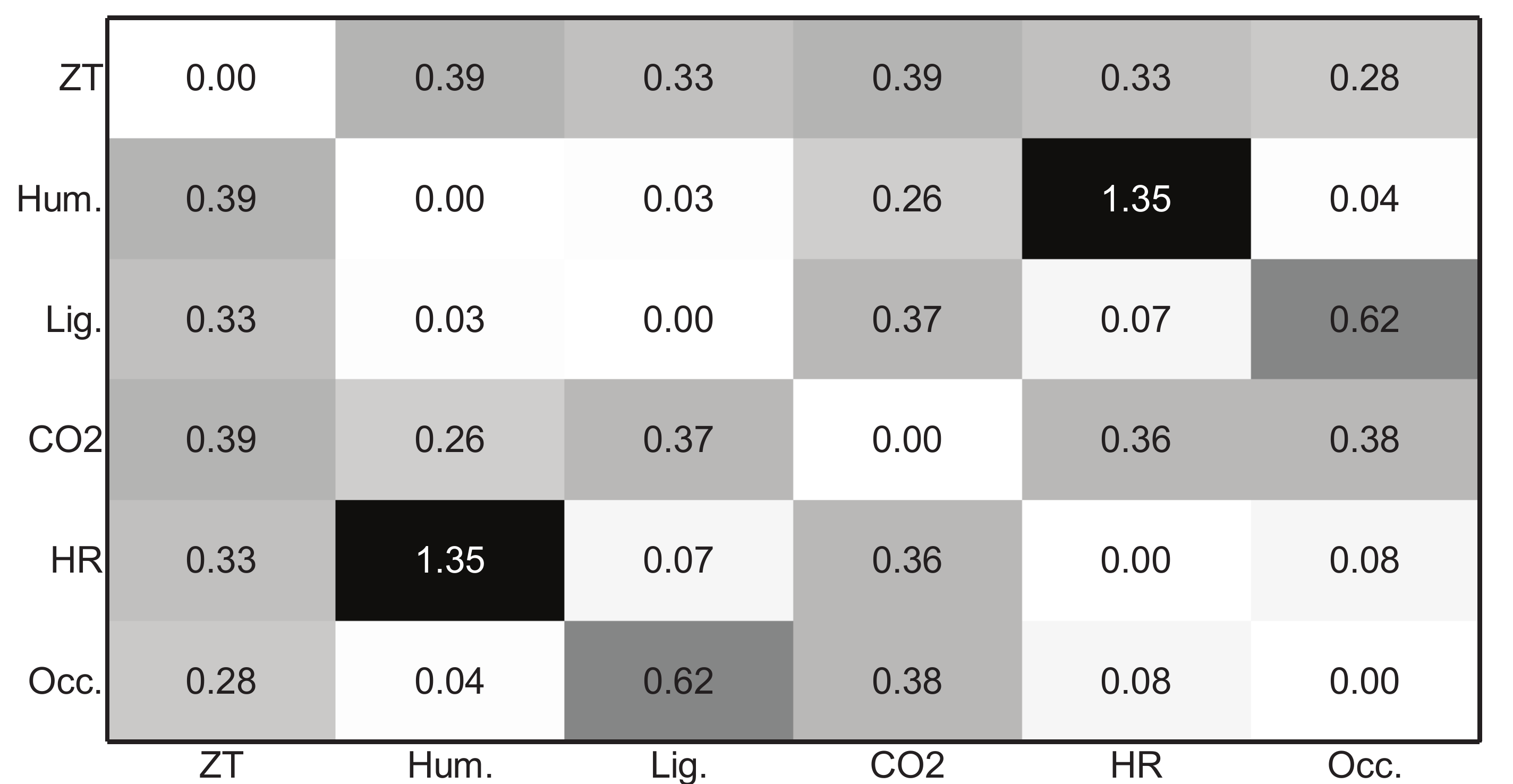}
  \caption{\textit{MI matrix - ZT: zone temperature; Hum.: humidity; Lig.: lighting; HR: humidity ratio; Occ.: occupancy}}\label{Figure0}
\end{figure}

We now use MI to calculate the causal dependencies for the above-mentioned dataset. 
The MI matrix can be obtained as shown in Figure~\ref{Figure0}.

As discussed before, only relational patterns are considered in this paper, such that entries along the diagonal of the MI-matrix is set to 0. The first column describes the causal dependencies of using other variables to predict the indoor air temperature. It can be seen that, from Figure~\ref{Figure0}, the calculated MI-value cannot match the prediction via ARMAX, as the MI-value of lighting (0.33), as a unique input, is smaller than that of humidity (0.39), as a unique input. Therefore, the information-theoretic metric based on MI does not reflect the true, underlying causal relationship. We introduce another information-theoretic metric based on {\it normalized} mutual information.

\begin{figure}
  \centering
  \vspace{3pt}
  \includegraphics[width=0.3\textwidth]{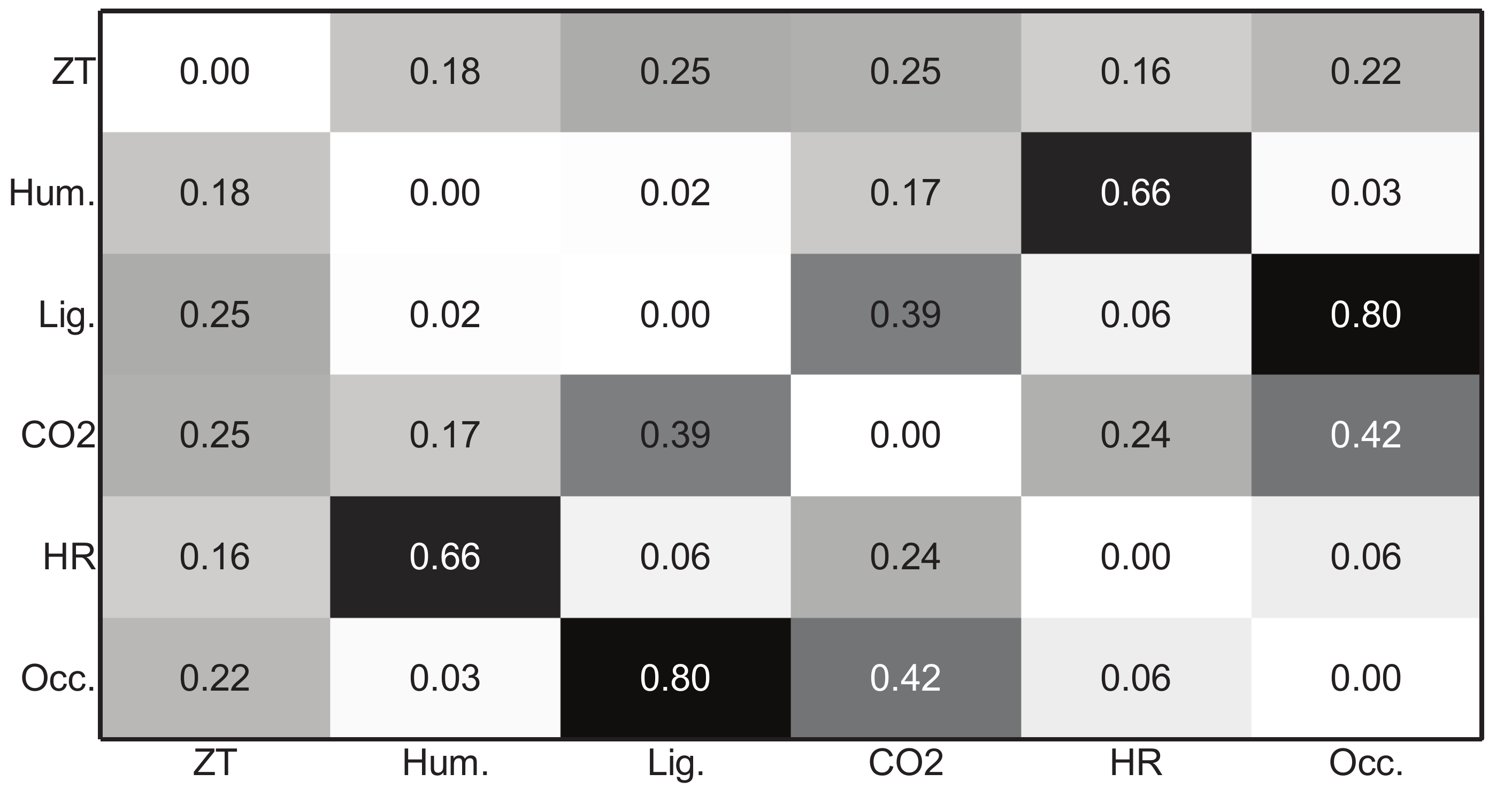}
  \caption{\textit{NMI matrix - ZT: zone temperature; Hum.: humidity; Lig.: lighting; HR: humidity ratio; Occ.: occupancy}}\label{Figure0_1}
\end{figure}

Normalized mutual information (NMI) is the normalization of MI, which has been proposed and shown to improve the sensitivity of MI, w.r.t. the difference in distribution, for two different random variables~\cite{vinh2010information}. Moreover, normalization can facilitate interpretation and comparison across different conditions, where measures might have different ranges. Different metrics based on NMI have been proposed, e.g., joint NMI~\cite{yao2003information}, maximum NMI~\cite{kvalseth1987entropy}, square-root NMI~\cite{strehl2002cluster}, and summation NMI~\cite{kvalseth1987entropy}. Square-root NMI can quantify the causal dependencies in the range $[0,1]$, making it analogous to the Pearson correlation coefficient, and can be calculated as follows:
\begin{equation}\label{nmi}
NMI_{sqrt} = \frac{I(U;V)}{\sqrt{H(U)H(V)}},
\end{equation}
where $H(U)$ and $H(V)$ are the entropies of $U$ and $V$, respectively.
Based on the $NMI_{sqrt}$, the matrix for the amount of NMI can be obtained in Figure~\ref{Figure0_1}.

As shown in Figure~\ref{Figure0_1}, entries in the first column show that lighting (0.25) has the most significant correlation with the indoor air temperature. While it can be observed that $CO_2$ (0.25) predicts the indoor air temperature better than occupancy (0.22), based on the matrix, this seems to violate the prediction results given by ARMAX. However, compared to MI, NMI can quite well imply the similar results suggested by ARMAX in a unified evaluation criterion. 
For further validating the effectiveness of NMI in finding causal relations between the indoor air temperature and other variables, three other case studies are presented in the next section. An algorithmic overview, for using NMI to detect the casual dependencies amongst variables and for using ARMAX in model inference, is given below.

\begin{algorithm}\label{NMI_ARMAX}
    \caption{Information Theoretic-based ARMAX}
    \SetKwInOut{Input}{Input}

    \Input{ARMAX model order, system delay, time-series data}
    \text{Decide inputs and outputs for ARMAX model}\\
    \text{Normalize the time-series data (optional)}\\
    \text{Convert continuous time-series data to symbol sequences}\\
    \text{Use Eq.~\ref{MI} to calculate MI for selected pairwise variables}\\
    \text{Calculate NMI using Eq.~\ref{nmi}}\\
    \text{Select significant variables as exogenous inputs in Eq.~\ref{armax}}\\
    \text{Use ARMAX for model inference}
\end{algorithm}


\section{Case Studies and Discussion}
\label{Case_studies}
\subsection{Three Case Studies}
In this section, we use three other indoor thermal environment datasets to validate our proposed approach. Furthermore, for comparison with other approaches, we make use of ARX, regularized ARMAX, and state-space models. The brief description about these three data sets are as follows.

The first dataset (SML 2010~\cite{zamora2014line}) is divided into training and testing datasets; the sampling frequency is 15 minutes. Variables include indoor temperature ($^\circ$C), $CO_2$ (ppm), humidity (\%), light (lux), sunlight in west/south/east facade (lux), sun irradiance ($W/m^2$), and outdoor temperature ($^\circ$C). The training data was collected during the period [13/03/2012 11:45 am, 11/04/2012 6:30 am], and the test data during [18/04/2012 12:00 am, 02/05/2012 07:30 am].

The second dataset (OpenEI~\cite{openei}) has a 1-minute sampling frequency. Variables selected are: zone temperature ($^\circ$F), air flow (CFM), discharge air static pressure, discharge air temperature ($^\circ$F), outside air temperature ($^\circ$F), and return air temperature ($^\circ$F). The collection period was [7/20/2014 00:01am, 4/1/2014 00:00 am]. This dataset is also divided into train/test datasets for the purpose of model-inference.

The third dataset (Iowa Interlock House~\cite{liu2016unsupervised,CBER}) has a 1-minute sampling frequency. Variables include: main floor temperature ($^\circ$C), outside air temperature ($^\circ$C), sunspace temperature ($^\circ$C), main floor relative humidity (\%), natural ventilation (mps), illuminance ($W/m^2$), and irradiance (lux). The data collection time was from 03/01/2014 to 04/01/2014.

Now, based on the proposed algorithm in the last section, we discuss these datasets by showing their NMI matrices and model-prediction capabilities. As we only consider the prediction of indoor air temperature using other variables, the first column is the focus in our discussion.

\begin{figure}
  \centering
  \includegraphics[width=0.3\textwidth]{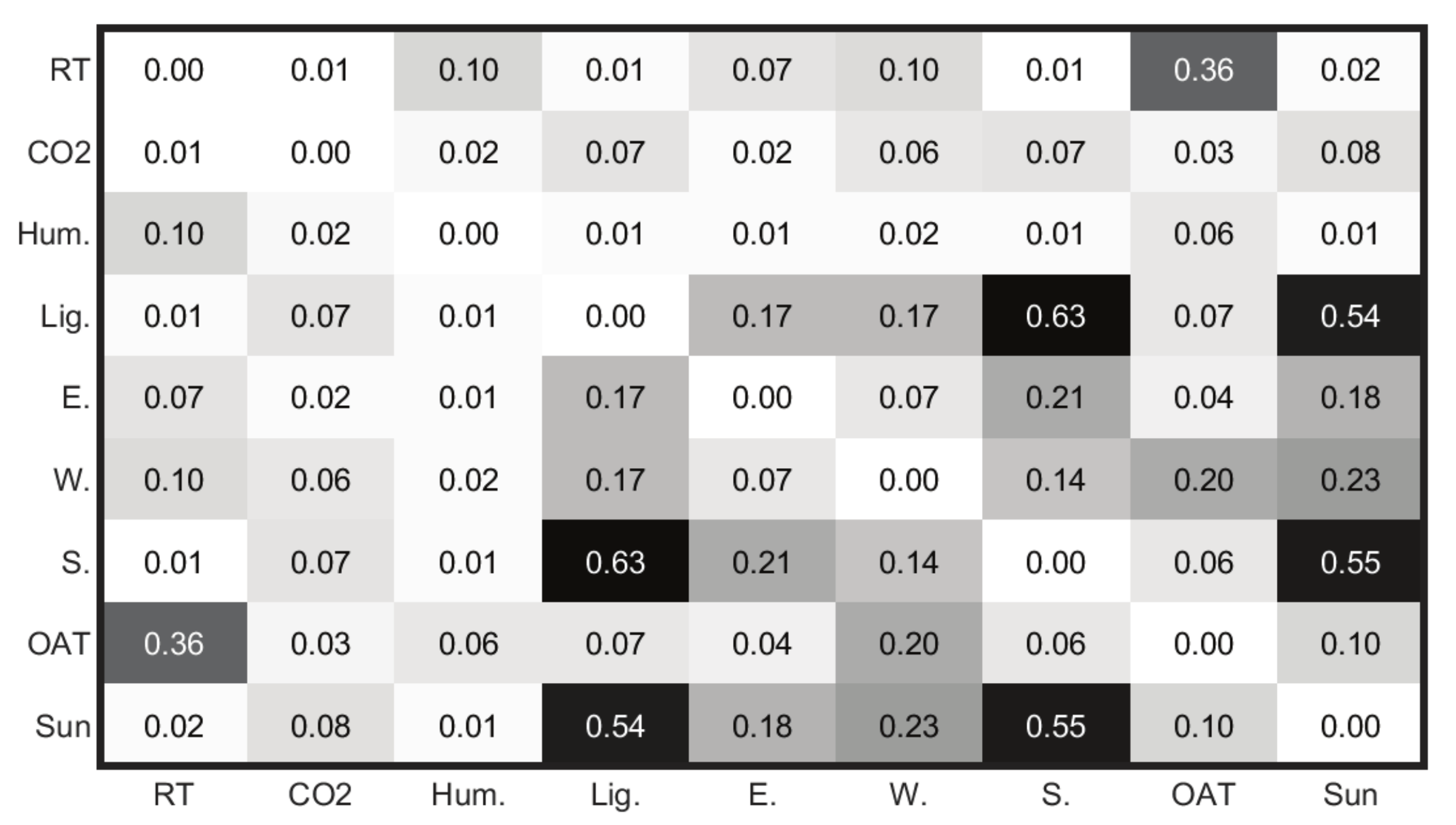}
  \caption{\textit{NMI matrix of the SML 2010 dataset - RT: room temperature; Hum.: humidity; Lig.: lighting; E.: sun light in east facade; W.: sun light in west facade; S.: sun light in south facade; OAT: outside air temperature; Sun: sun irradiance}}\label{dataset2}
\end{figure}

\begin{figure}
  \centering
  \includegraphics[width=0.35\textwidth]{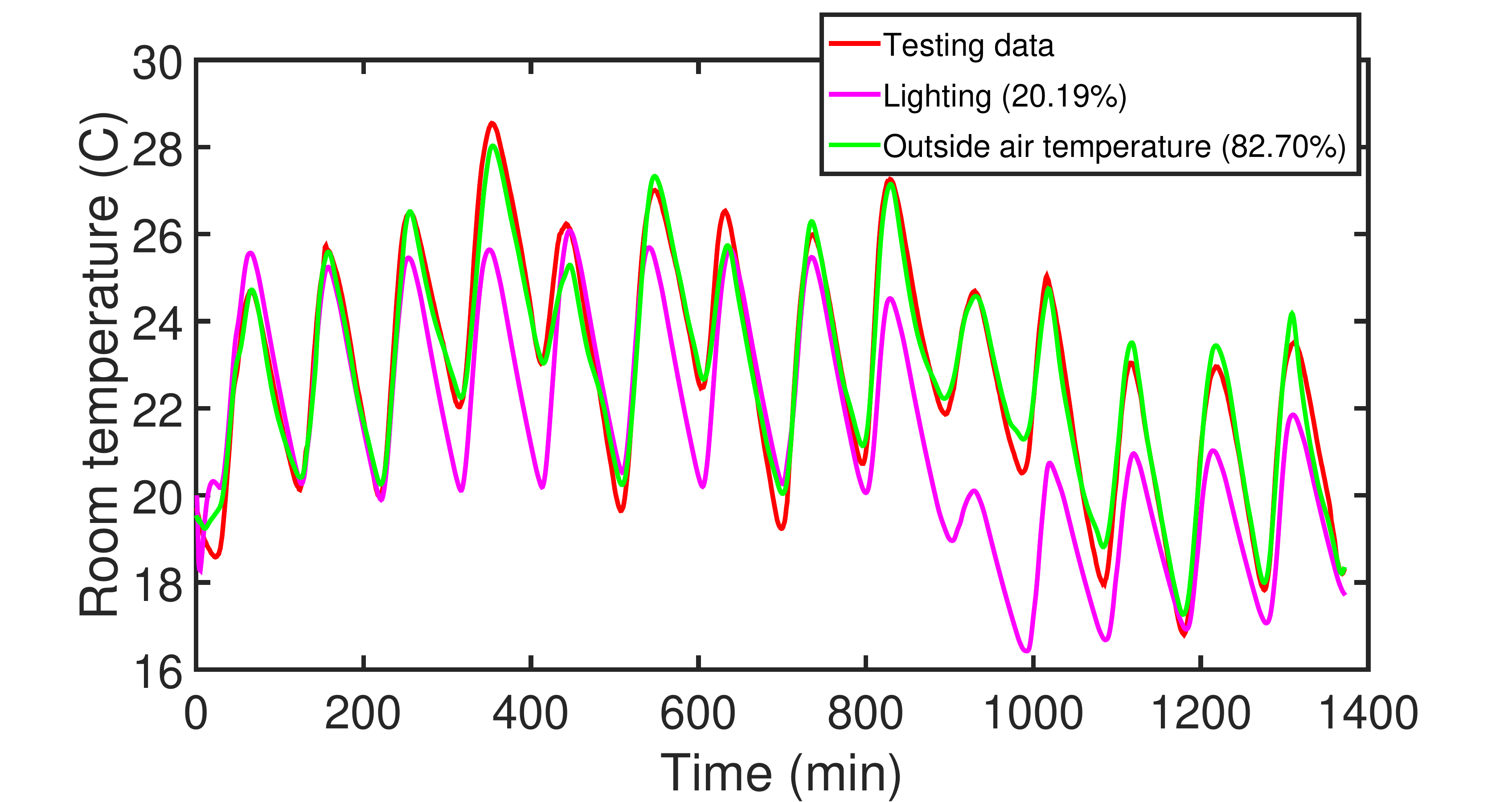}
  \caption{\textit{SISO with SML 2010 dataset}}\label{Figure8}
\end{figure}

\textit{First dataset}: The NMI matrix is as shown in Figure~\ref{dataset2}.
The first column shows that the model predicting capability, based on the outside air temperature as the input, is better than any other variable. Its NMI is up to 0.36. Therefore, we select lighting (0.01) and outside air temperature as the exogenous input for the ARMAX model for comparison. Figure~\ref{Figure8} shows that, based on outside air temperature, the model fit is 82.70\% whereas it is 20.19\% for lighting. This implies that outside air temperature dominates other variables if selected as the exogenous input, which validates the suggestion obtained from the NMI matrix.

\begin{figure}
  \centering
  \includegraphics[width=0.3\textwidth]{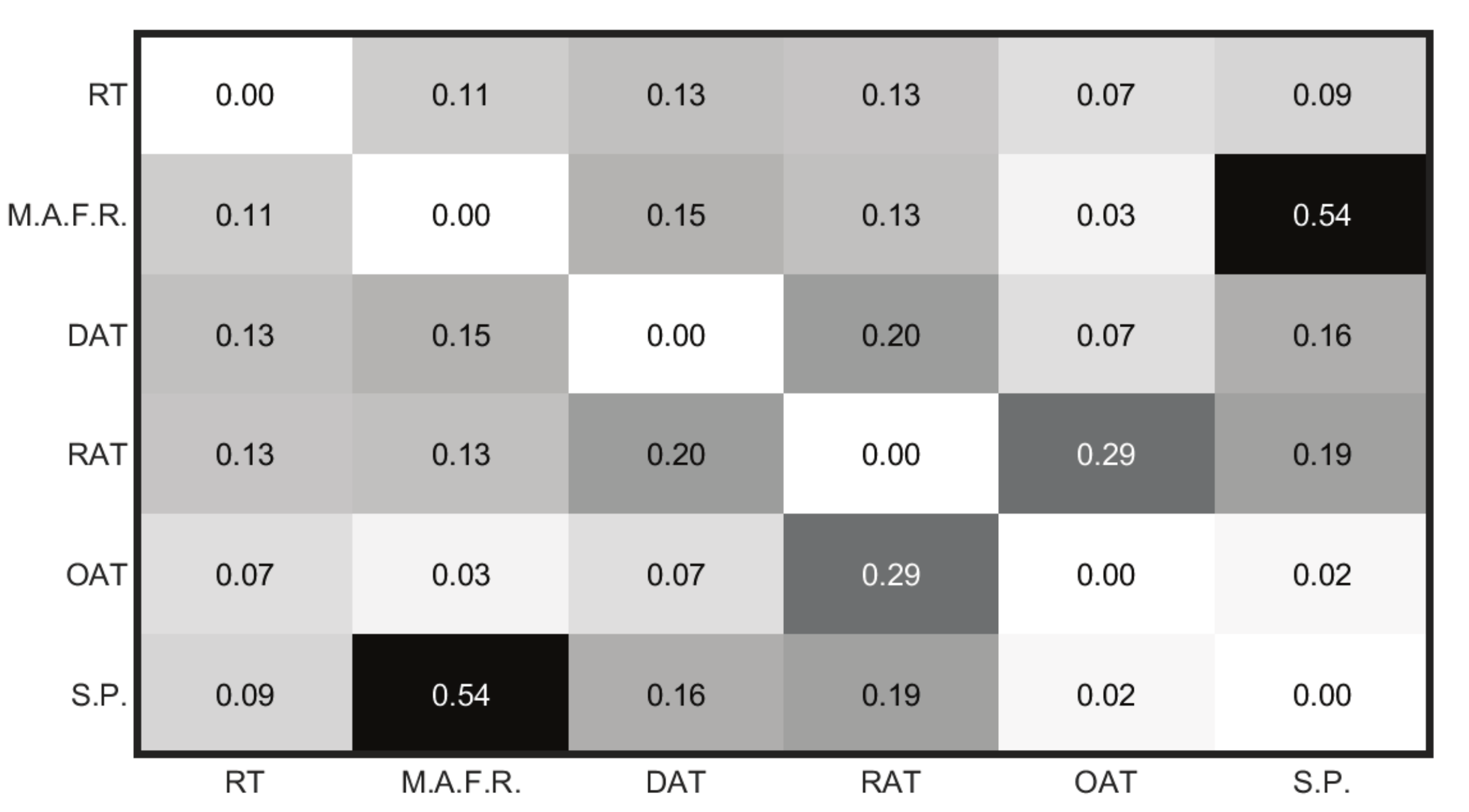}
  \caption{\textit{NMI matrix of the OpenEI dataset - RT: room temperature; M.A.F.R.: air mass flow rate; DAT: discharge air temperature; RAT: return air temperature; OAT: outside air temperature; S.P.: static pressure of discharge air}}\label{dataset3}
\end{figure}

\begin{figure}
  \centering
  \vspace{5pt}
  \includegraphics[width=0.35\textwidth]{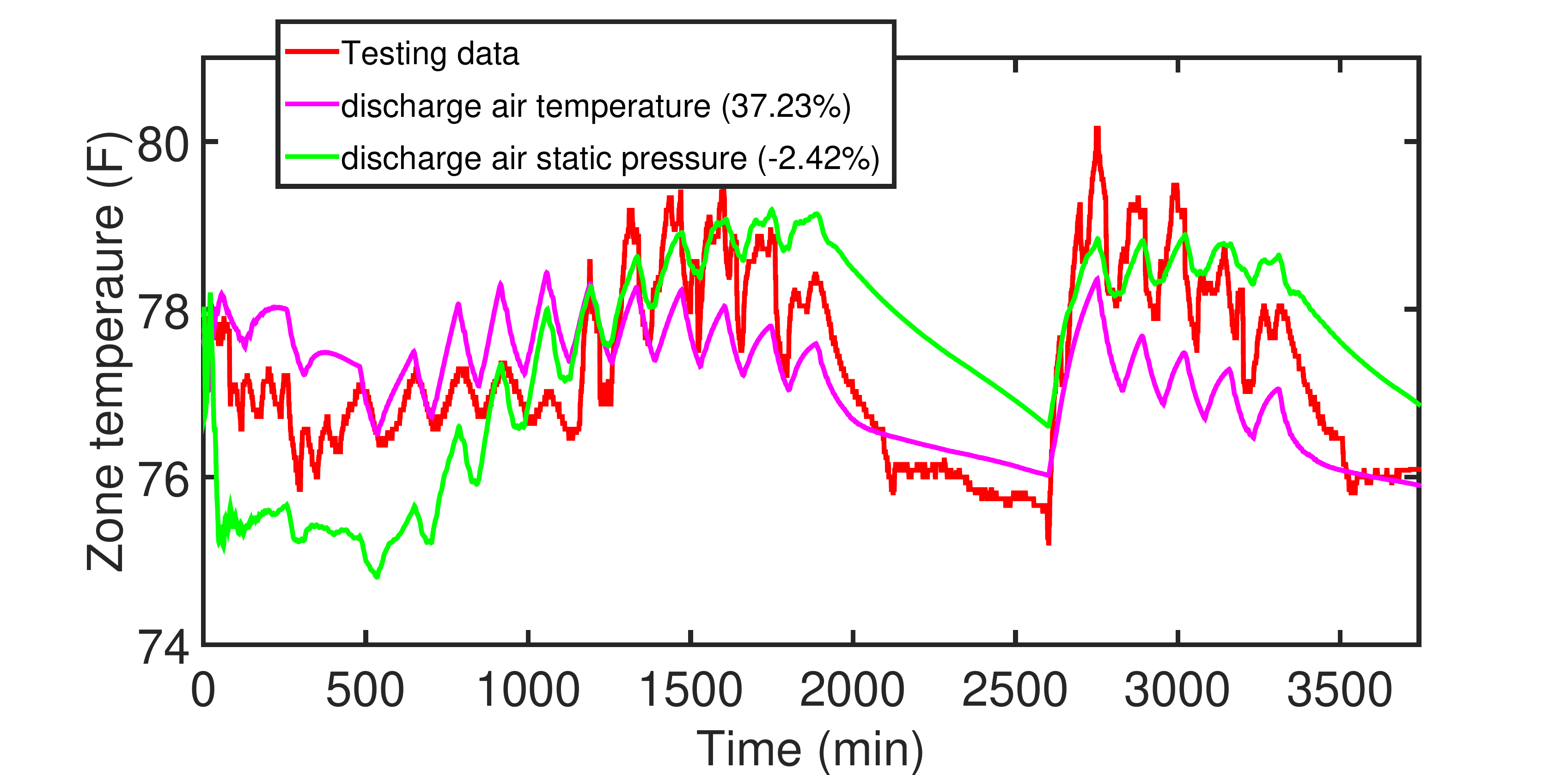}
  \caption{\textit{MISO with OpenEI dataset}}\label{Figure9}
\end{figure}

\textit{Second dataset}: The NMI matrix is as shown in Figure~\ref{dataset3}.
As the entry values are small|which implies that the model prediction capability using only one variable as the exogenous input may not be quite good|we use MISO to show the prediction. We select discharge air temperature (0.13) and discharge air static pressure (0.09) for tested variables and select air mass flow rate (0.11) and return air temperature (0.13) as auxiliary variables. Therefore, this case is with 3 inputs and 1 output. The results on the model fit are shown in Figure~\ref{Figure9}.
One conclusion that can be made is that the variable with highest NMI can improve the model prediction capability most significantly with 37.23\% model fit. 

\begin{figure}
  \centering
  \includegraphics[width=0.35\textwidth]{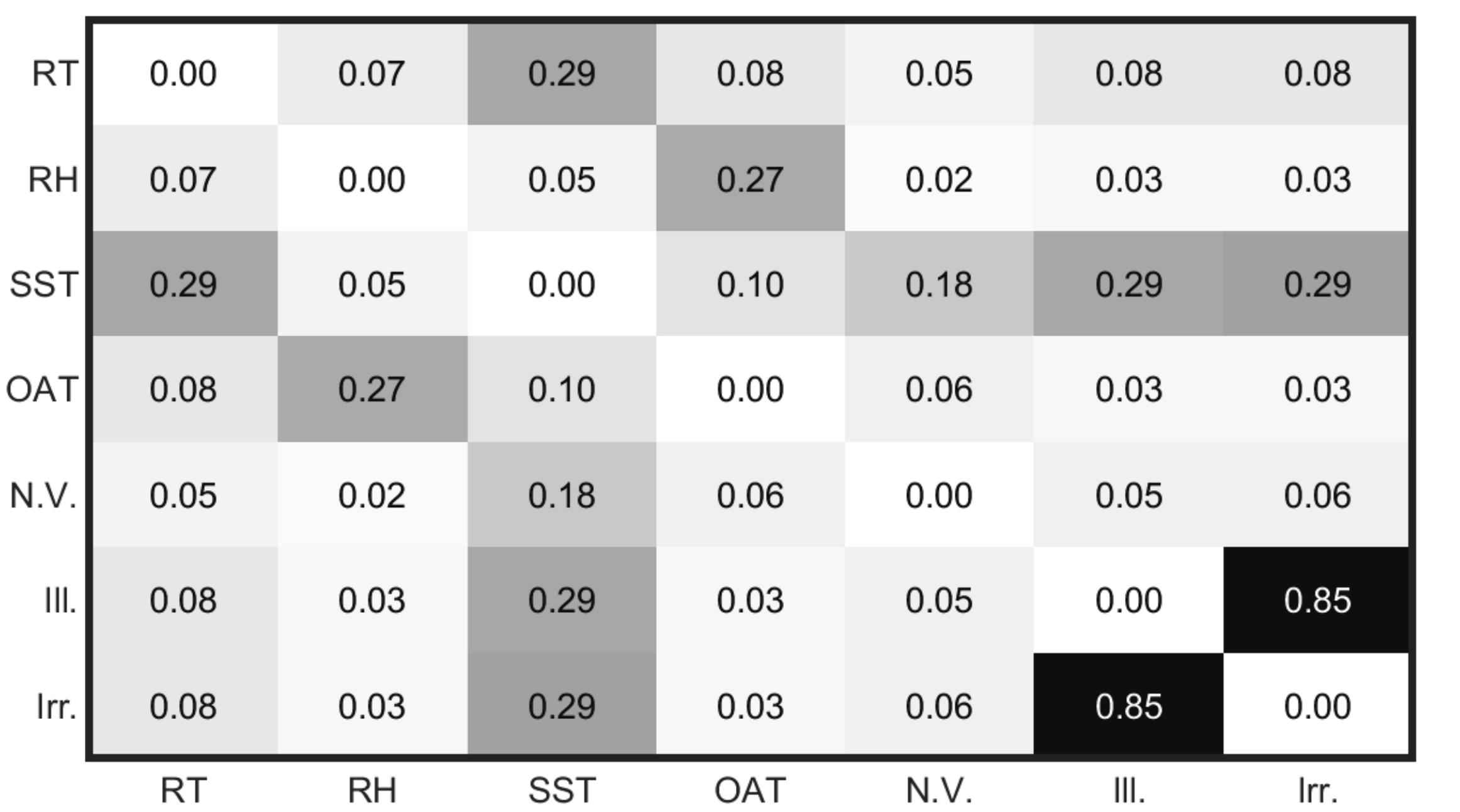}
  \caption{\textit{NMI matrix of the Iowa Interlock House dataset - RT: room temperature; RH: relative humidity; SST: sunspace temperature; OAT: outside air temperature; N.V.: natural ventilation; Ill.: sun illuminance; Irr: sun irradiance}}\label{dataset4}
\end{figure}

\textit{Third dataset}: The NMI matrix is shown in Figure~\ref{dataset4}.
Similarly, we choose the variable with highest NMI value as the unique exogenous input, i.e., sunspace temperature, and compare the model predicting capability with that of another variable, natural ventilation (0.05).
\begin{figure}
  \centering
  \includegraphics[width=0.35\textwidth]{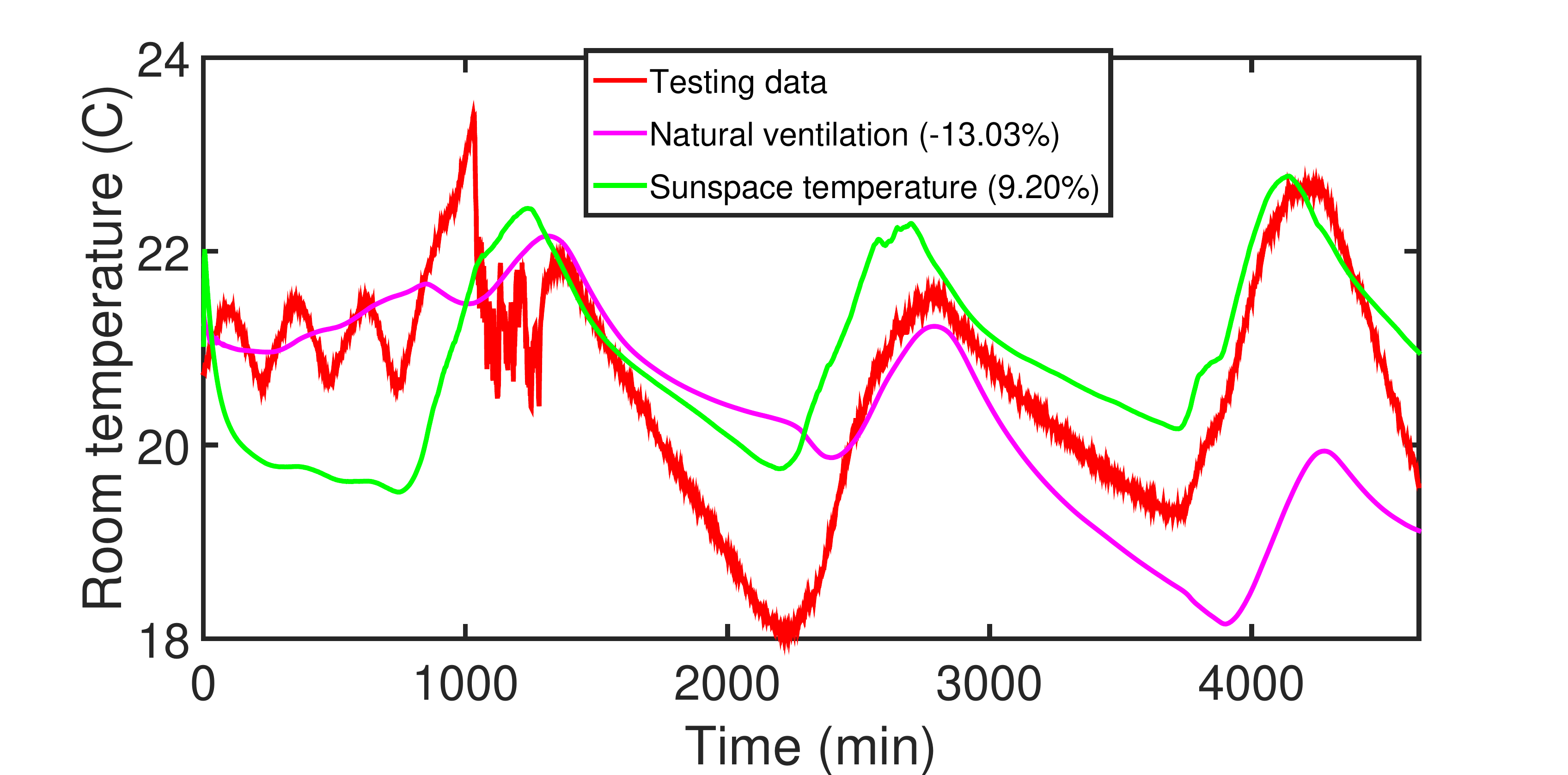}
  \caption{\textit{SISO with the Iowa Interlock House dataset}}\label{Figure10}
\end{figure}
As shown in Figure~\ref{Figure10}, a single exogenous input cannot predict the indoor air temperature well. However, it turns out that sunspace temperature has more impact on indoor air temperature, which can be implied by the house configuration. Although natural ventilation can follow the trend of the indoor air temperature, it cannot perform well on the temperature peaks and valleys.

\subsection{Comparison with ARX, Regularized ARMAX, and State-space Models}
This section presents the comparison on model predicting capability with ARX, regularized ARMAX, and state-space models using the dataset in the illustrative example. 
The dataset used for the purpose of comparison is the dataset used in the illustrative example in section~\ref{ARMAX}.
\begin{figure}
  \centering
  \vspace{5pt}
  \includegraphics[width=0.35\textwidth]{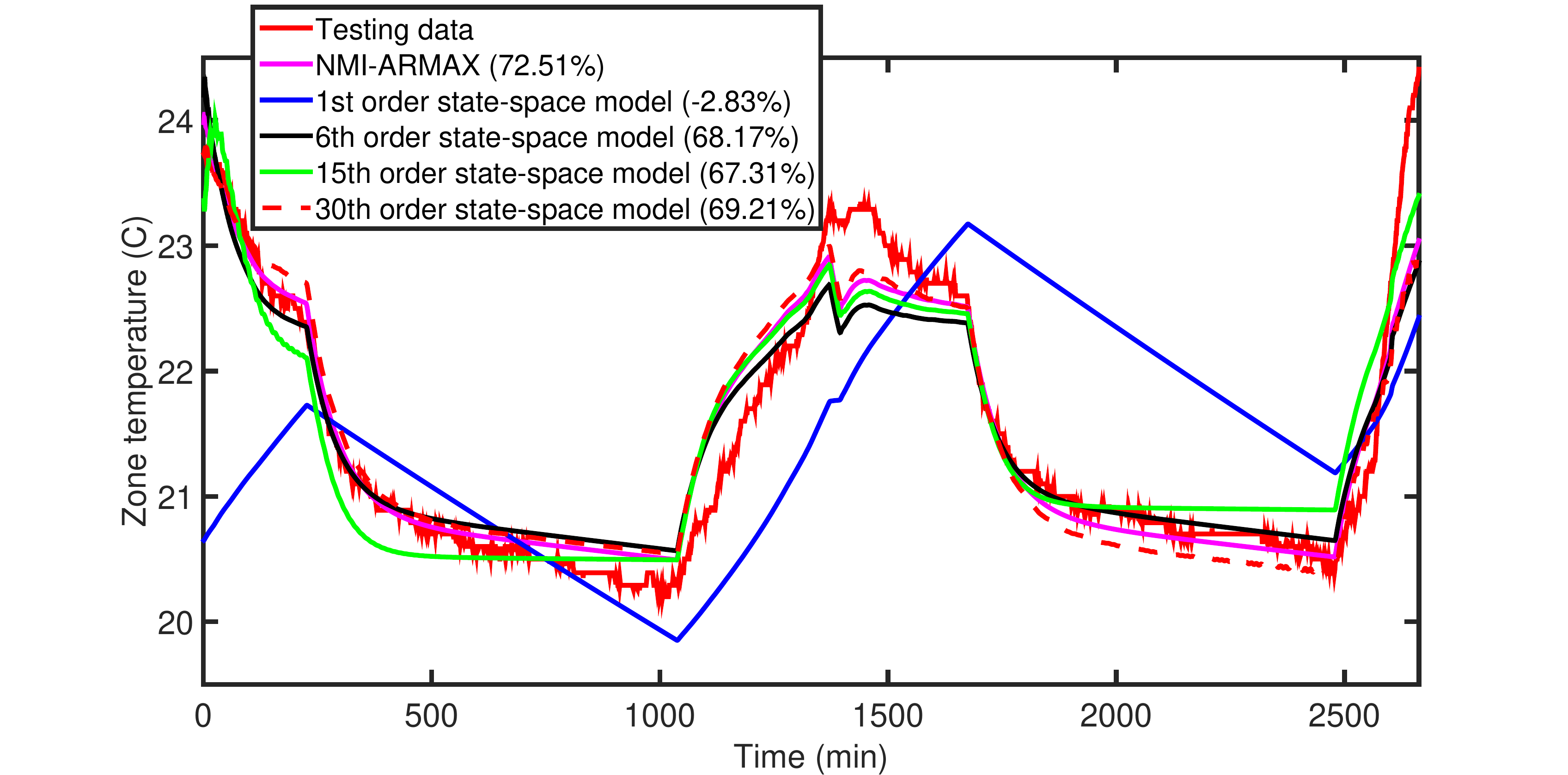}\vspace{-8pt}
  \caption{\textit{Comparison between ARMAX and state-space models for the first testing data set}}\label{Figure11}
\end{figure}

\begin{figure}
  \centering
  \includegraphics[width=0.35\textwidth]{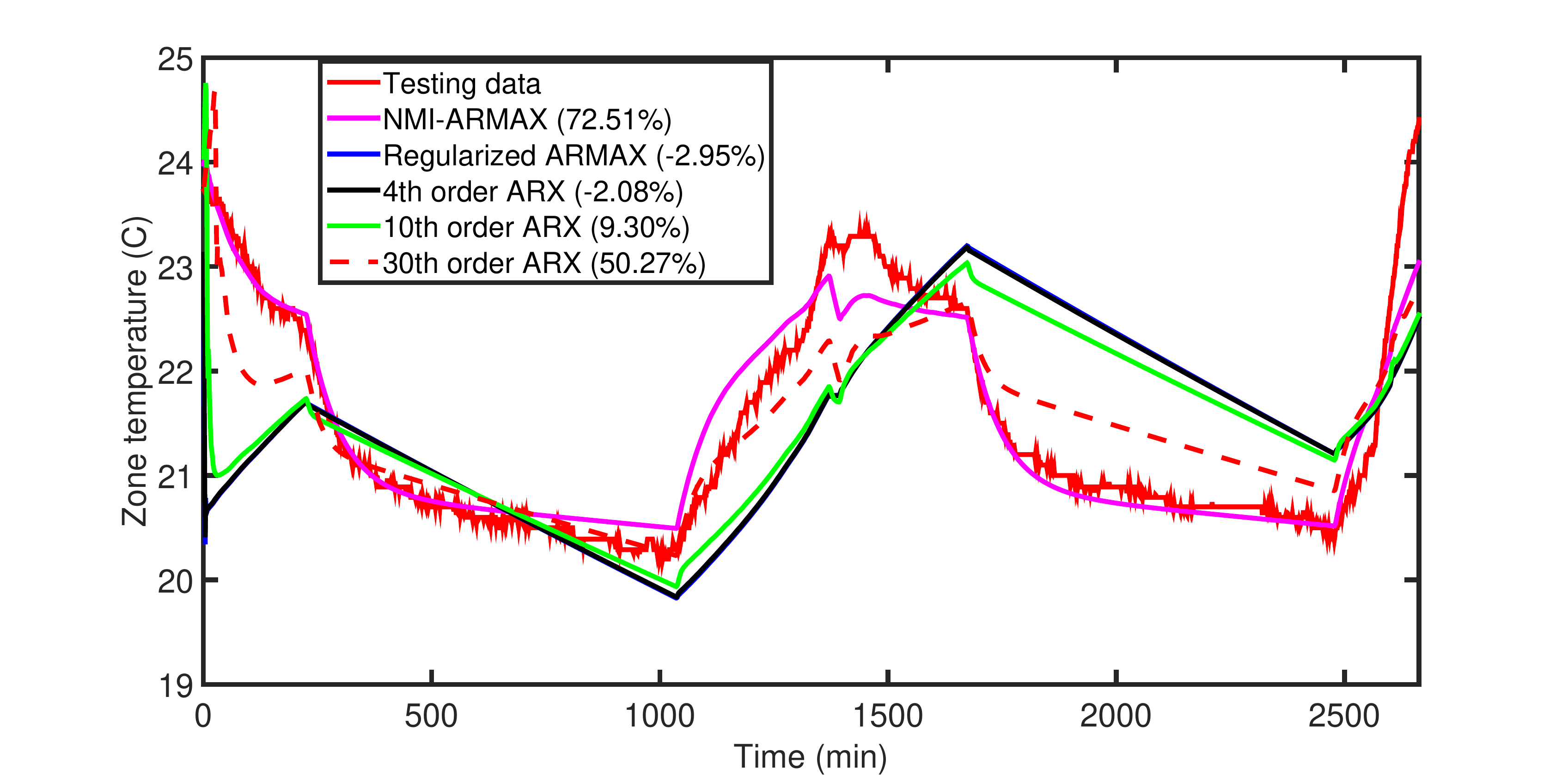}\vspace{-8pt}
  \caption{\textit{Comparison between ARMAX and ARX and regularized ARMAX models for the first testing data set}}\label{Figure12}
\end{figure}

\begin{figure}
  \centering
  \includegraphics[width=0.35\textwidth]{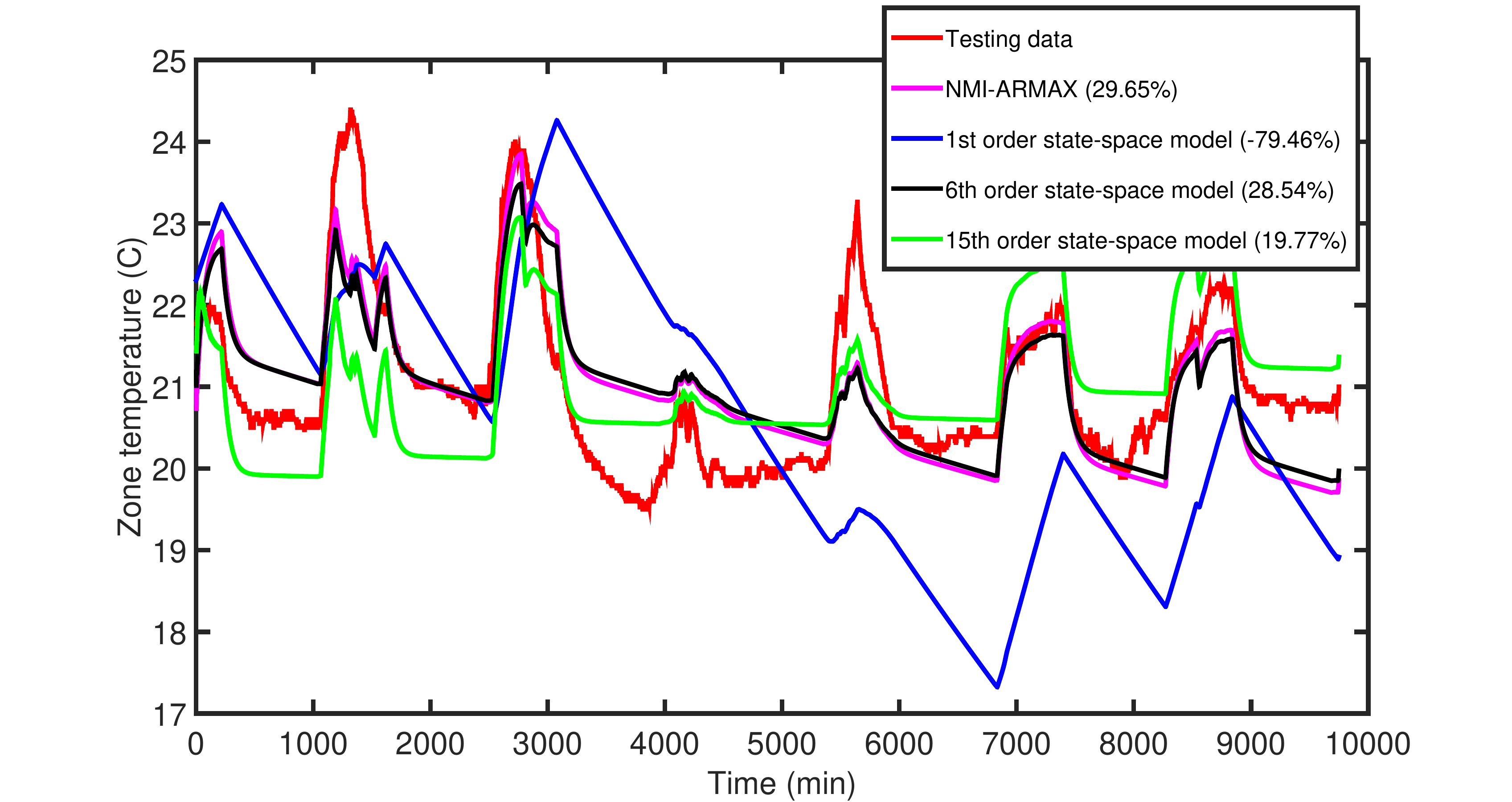}\vspace{-10pt}
  \caption{\textit{Comparison between ARMAX and state-space models for the second testing data set}}\label{Figure13}
\end{figure}

\begin{figure}
  \centering
  \includegraphics[width=0.35\textwidth]{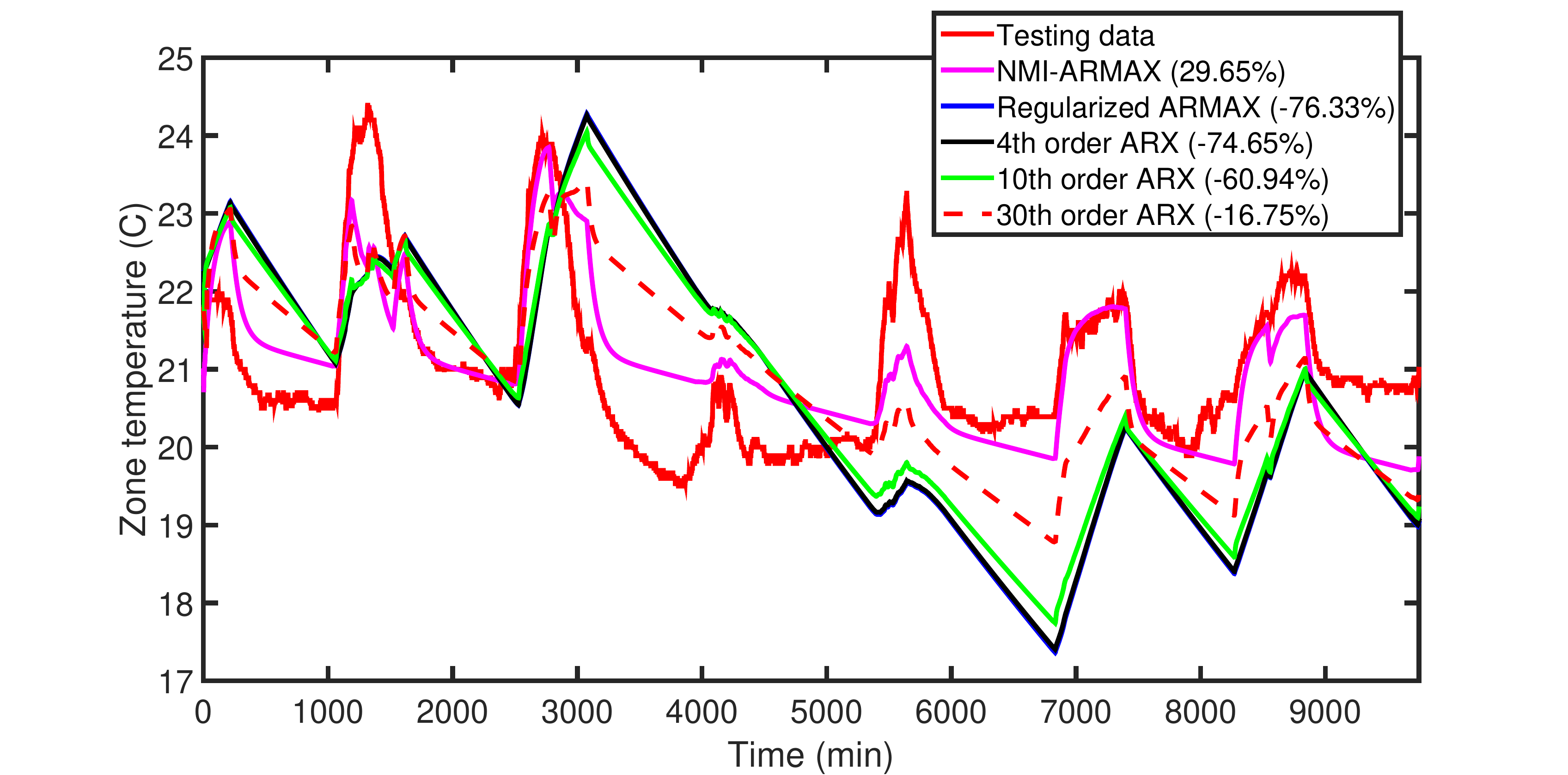}\vspace{-10pt}
  \caption{\textit{Comparison between ARMAX and ARX and regularized ARMAX models for the second testing data set}}\label{Figure14}
\end{figure}

We first discuss the SISO case in Figures~\ref{Figure11} -~\ref{Figure14} for two testing datasets, respectively. Figures~\ref{Figure11} and~\ref{Figure13} show the comparison between ARMAX and state-space models, with different orders (1st, 6th, 15th, and 30th). The results show that ARMAX of lower orders can out-perform state-space models. Only when the state-space model is 15th order is the model-fit comparable to that of ARMAX. It is noted that 30th order model fit in Figure~\ref{Figure13} is not shown in this context as the result is significantly non-comparable. Table~\ref{Table2} shows model fit results using different schemes with three inputs. In Table~\ref{Table2}, S.S. represents state-space and Reg. is regularized. It can be observed that only when the state-space models are, respectively, 15th and 30th orders, is the model fit comparable to that of ARMAX. Implicitly, for state-space models, higher orders are required to obtain higher model accuracy. However, in practice, lower-order state-space models are more feasibly used; the higher the model order is, the higher the number of states required. This correspondingly increases the computational difficulty and complexity of controller design. Furthermore, as shown in Figures~\ref{Figure12} and~\ref{Figure14}, ARMAX with single input (lighting suggested by NMI) has better model fit compared to regularized ARMAX and ARX. In Table~\ref{Table2}, even with multiple inputs, ARMAX still outperforms ARX and regularized ARMAX. For ARX, it can be concluded similarly that higher orders can improve the model accuracy at the cost of computational difficulty and complexity. In this case, however, regularization negatively affects the model predicting capability of ARMAX due to the reduction of model fit. For example, when the data set is the first testing data set, regularized ARMAX reduces the model fit by 46.29\%, compared to ARMAX. It is because the regularization introduces the bias for the model. Therefore, regularization for ARMAX should be chosen more carefully in indoor thermal model inference.

\begin{table*}
\centering
\vspace{4pt}
\caption{Model fit using different methods for three inputs (lighting + occupancy + $CO_2$)}
\begin{tabular}{|l|*{9}{c|}}\hline
\backslashbox{Data set}{Method}
&ARMAX&1st S.S.&6th S.S.&15th S.S.&30th S.S.& Reg. ARMAX&4th ARX&10th ARX&30th ARX
\\\hline
First testing data set &\textbf{73.96\%}&24.80\%&60.32\%&72.13\%&77.79\%&27.67\%&28.71\%&41.83\%&61.93\%\\\hline
Second testing data set & \textbf{43.46\%}&-281.27\%&-43.37\%&13.21\%&36.33\%&-263.44\%&-255.64\%&-154.74\%&-22.13\%\\\hline
\end{tabular}
\label{Table2}
\end{table*}


\section{Conclusions and Future Work}
\label{conlusions}

This paper proposes data-driven indoor thermal model inference, with the combination of ARMAX and information-theoretic methods. By using NMI, the causal dependency (quantified by the information value between the output variable and any other variable) can be obtained as a selection-guideline of exogenous inputs for ARMAX. Three case-studies based on building thermal environments were presented for validation. Comparison with state-space, ARX, and regularized ARMAX models is conducted to demonstrate the efficacy of ARMAX. Our proposed framework can help reduce model complexity and computation resources in coefficient-estimation, while maintaining model accuracy. Beyond the existing work, some future directions include: 1) using probabilistic graphical models (e.g., Bayesian Networks) for indoor thermal dynamics; 2) incorporating the proposed method into model predictive controller design.



\section*{ACKNOWLEDGMENT}

This paper is based upon research partially supported by the U.S. Department of Energy grant DE-EE0007682. The Interlock House data collection is managed by the Center for Building Energy Research at Iowa State University and supported by the National Science Foundation Grant Number EPSC-1101284.  Any opinions, findings, and conclusions or recommendations expressed in this material are those of the author(s) and do not necessarily reflect the views of the National Science Foundation or the U.S. Department of Energy.


\bibliographystyle{ieeetr}
\bibliography{ARMAX_NMI}

\begin{thebibliography}{10}

\bibitem{enescu2017review}
D.~Enescu, ``A review of thermal comfort models and indicators for indoor
  environments,'' {\em Renewable and Sustainable Energy Reviews}, vol.~79,
  pp.~1353--1379, 2017.

\bibitem{kalogirou2000applications}
S.~A. Kalogirou, ``Applications of artificial neural-networks for energy
  systems,'' {\em Applied energy}, vol.~67, no.~1, pp.~17--35, 2000.

\bibitem{rios2007modelling}
G.~R{\'\i}os-Moreno, M.~Trejo-Perea, R.~Castaneda-Miranda,
  V.~Hern{\'a}ndez-Guzm{\'a}n, and G.~Herrera-Ruiz, ``Modelling temperature in
  intelligent buildings by means of autoregressive models,'' {\em Automation in
  Construction}, vol.~16, no.~5, pp.~713--722, 2007.

\bibitem{chinde2015comparative}
V.~Chinde, J.~C. Heylmun, A.~Kohl, Z.~Jiang, S.~Sarkar, and A.~Kelkar,
  ``Comparative evaluation of control-oriented zone temperature prediction
  modeling strategies in buildings,'' in {\em ASME 2015 Dynamic Systems and
  Control Conference}, pp.~V002T34A009--V002T34A009, American Society of
  Mechanical Engineers, 2015.

\bibitem{marvuglia2014coupling}
A.~Marvuglia, A.~Messineo, and G.~Nicolosi, ``Coupling a neural network
  temperature predictor and a fuzzy logic controller to perform thermal comfort
  regulation in an office building,'' {\em Building and Environment}, vol.~72,
  pp.~287--299, 2014.

\bibitem{liang2005thermal}
J.~Liang and R.~Du, ``Thermal comfort control based on neural network for hvac
  application,'' in {\em Control Applications, 2005. CCA 2005. Proceedings of
  2005 IEEE Conference on}, pp.~819--824, IEEE, 2005.

\bibitem{moon2016algorithm}
J.~W. Moon and S.~K. Jung, ``Algorithm for optimal application of the setback
  moment in the heating season using an artificial neural network model,'' {\em
  Energy and Buildings}, vol.~127, pp.~859--869, 2016.

\bibitem{moon2010ann}
J.~W. Moon and J.-J. Kim, ``Ann-based thermal control models for residential
  buildings,'' {\em Building and Environment}, vol.~45, no.~7, pp.~1612--1625,
  2010.

\bibitem{mustafaraj2010development}
G.~Mustafaraj, J.~Chen, and G.~Lowry, ``Development of room temperature and
  relative humidity linear parametric models for an open office using bms
  data,'' {\em Energy and Buildings}, vol.~42, no.~3, pp.~348--356, 2010.

\bibitem{wu2012physics}
S.~Wu and J.-Q. Sun, ``A physics-based linear parametric model of room
  temperature in office buildings,'' {\em Building and Environment}, vol.~50,
  pp.~1--9, 2012.

\bibitem{mechaqrane2004comparison}
A.~Mechaqrane and M.~Zouak, ``A comparison of linear and neural network arx
  models applied to a prediction of the indoor temperature of a building,''
  {\em Neural Computing \& Applications}, vol.~13, no.~1, pp.~32--37, 2004.

\bibitem{candanedo2016accurate}
L.~M. Candanedo and V.~Feldheim, ``Accurate occupancy detection of an office
  room from light, temperature, humidity and co 2 measurements using
  statistical learning models,'' {\em Energy and Buildings}, vol.~112,
  pp.~28--39, 2016.

\bibitem{jiang2017energy}
Z.~Jiang, C.~Liu, A.~Akintayo, G.~Henze, and S.~Sarkar, ``Energy prediction
  using spatiotemporal pattern networks,'' {\em arXiv preprint
  arXiv:1702.01125}, 2017.

\bibitem{vinh2010information}
N.~X. Vinh, J.~Epps, and J.~Bailey, ``Information theoretic measures for
  clusterings comparison: Variants, properties, normalization and correction
  for chance,'' {\em Journal of Machine Learning Research}, vol.~11, no.~Oct,
  pp.~2837--2854, 2010.

\bibitem{yao2003information}
Y.~Yao, ``Information-theoretic measures for knowledge discovery and data
  mining,'' in {\em Entropy Measures, Maximum Entropy Principle and Emerging
  Applications}, pp.~115--136, Springer, 2003.

\bibitem{kvalseth1987entropy}
T.~O. Kvalseth, ``Entropy and correlation: Some comments,'' {\em IEEE
  Transactions on Systems, Man, and Cybernetics}, vol.~17, no.~3, pp.~517--519,
  1987.

\bibitem{strehl2002cluster}
A.~Strehl and J.~Ghosh, ``Cluster ensembles---a knowledge reuse framework for
  combining multiple partitions,'' {\em Journal of machine learning research},
  vol.~3, no.~Dec, pp.~583--617, 2002.

\bibitem{zamora2014line}
F.~Zamora-Mart{\'\i}nez, P.~Romeu, P.~Botella-Rocamora, and J.~Pardo, ``On-line
  learning of indoor temperature forecasting models towards energy
  efficiency,'' {\em Energy and Buildings}, vol.~83, pp.~162--172, 2014.

\bibitem{openei}
OpenEI, ``Long-term data on 3 office air handling units.''
  \url{https://en.openei.org/datasets/dataset/long-term-data-on-3-office-air-handling-units},
  2017-07-04.

\bibitem{liu2016unsupervised}
C.~Liu, S.~Ghosal, Z.~Jiang, and S.~Sarkar, ``An unsupervised spatiotemporal
  graphical modeling approach to anomaly detection in distributed cps,'' in
  {\em Cyber-Physical Systems (ICCPS), 2016 ACM/IEEE 7th International
  Conference on}, pp.~1--10, IEEE, 2016.

\bibitem{CBER}
C.~of~Building Energy~Research, ``Interlock house.''
  \url{http://www.cber.iastate.edu/projects/interlock-house}, 2013.

\end{thebibliography}

\end{document}